\documentclass[12pt]{amsart}
\usepackage{amsmath,amssymb,amsfonts}
\usepackage{epsfig}
\usepackage[english]{babel}
\usepackage{graphics}
\newcommand{\bs}{\boldsymbol}
\newcommand{\cC}{\mathcal{C}}
\newcommand{\cI}{\mathcal{I}}
\newcommand{\cA}{\mathcal {A}}

\def\half{\mbox{\small $\frac{1}{2}$}}
\def\thalf{\mbox{ $\scriptstyle\frac{1}{2}$}}

\begin{document}
\begin{titlepage}
  \begin{flushright}
  \end{flushright}
\vskip 1.in
  \begin{center}
{\large Landau levels on the 2-D torus:
  a numerical approach} 
\par \vskip 5mm 
{\sf Enrico Onofri} $^{a, b}$
\end{center}

\par \vskip 10mm
\begin{center} {\bf Abstract} 
 \end{center}
 \begin{quote} {\sl A numerical method is presented which allows to
     compute the spectrum of the Schroedinger operator for a particle
     constrained on a two dimensional flat torus under the combined
     action of a transverse magnetic field and any conservative
     force. The method employs a fast Fourier transform to accurately
     represent the momentum variables and takes into account the
     twisted boundary conditions required by the presence of the
     magnetic field. An accuracy of twelve digits is attained even
     with coarse grids. Landau levels are reproduced in the case of a
     uniform field satisfying Dirac's condition. A new fine structure
     of levels within the single Landau level is formed when the field
     has a sinusoidal component with period commensurable to the
     integer magnetic charge.}
\end{quote}

\vfill
\begin{flushleft}
  {\sf PACS numbers: 31.15.-p, 71.70.Di}
\end{flushleft}
\vskip .2 in

\noindent { $^{a)}$ \small\sl Laboratoire de Physique
  Th\'eorique et Astroparticules, Universit\'e Montpellier II, Place
  E. Bataillon, 34095 Montpellier Cedex 05, France. }

\noindent
{$^{b)}$ \small\sl Permanent address: Universit\`a di Parma and
  INFN, Gruppo Collegato di Parma, 43100 Parma, Italy.}

\keywords{Landau Levels, monopole, Dirac's quantization}


\end{titlepage}
\section{Introduction}\label{sec:introduction}
The quantum mechanics of a charged particle living on a
two-dimen\-sional torus in presence of a uniform magnetic field,
orthogonal to the surface, has been solved years ago \cite{fubini92,
  onofri01, morariu01}. The degeneration of the ground state coincides
with the flux of the magnetic field, in units of the elementary flux $
h c/e $ (in this paper we shall adopt units such that
$\hslash=e/c=1$). This is a simple example of the more general theorem
about cohomology groups for hermitian line bundles
\cite{hirzebruch78}, known in the physical literature as Dirac's
quantization condition: quantum mechanics requires that the flux of
the magnetic field across a closed surface must be quantized. This is
also known as the Weil-Souriau-Kostant quantization condition.

In this paper we present a numerical algorithm which is accurate
enough in representing the momentum variables and it respects the
constraints posed by differential geometry. The algorithm computes the
spectrum of the quantum particle on the torus in presence of both a
transverse magnetic field and a scalar potential. If the potential
vanishes and the magnetic field is uniform the algorithm reproduces
the known spectrum, in terms of eigenvalues and degeneration, to a
typical accuracy of twelve digits. The effect of the potential energy
is to split the Landau Levels; this fact is at the basis of Klauder's
formulation of path integrals in phase space \cite{Klauder} and our
algorithm could be used to explore this approach to quantization
theory. The case of a non-uniform magnetic field and the corresponding
splitting pattern of Landau levels can be studied using our
algorithm. We consider the case of a sinusoidal contribution to the
magnetic field in the last section. A peculiar fine structure emerges,
which is made visible by the accuracy of the algorithm. This
fine-structure within each Landau level could be dubbed {\sl
  Landau-Mathieu levels} and it manifests itself when the number of
oscillations of the perturbed field is commensurate to the quantized
magnetic flux. This fact suggests that an undulating stationary
magnetic field could be used to tune the number of states in the fine
structure of Landau-Mathieu sub levels.

\section{The model}

Quantum mechanics on a compact surface, in the presence of a magnetic
field transverse to the surface, requires the introduction of either a
singular magnetic potential (Dirac's string) or a collection of local
potentials $A_\alpha$, one for each local chart of a given atlas on
the surface. The description in terms of local potentials is
preferable for its mathematical rigor \cite{alvarez85}. The
implementation of the local description within a numerical approach
should be easily achieved in terms of finite elements methods.  In
this paper we take an alternative route, working on a single chart,
but imposing the correct (twisted) boundary conditions to the wave
function, as we explain in the next section.

\section{Local charts and twisted boundary conditions}\label{sec:local-charts-twisted}
Let the torus be identified with the two-dimensional plane
$\mathbb{R}^2$ modulo the discrete subgroup of translations generated
by $x\to x+L_1, y\to y+L_2$.  We cover the torus with four charts
defined as follows
\begin{equation}
\begin{matrix}
\cC_\alpha:
\begin{cases}
  0<x<L_1\phantom{+\delta_2}\\
  0<y<L_2
\end{cases} &
\cC_\beta:
\begin{cases}
  \delta_1<x< L_1+\delta_1\\
  0<y<L_2
\end{cases}\\[.2em]
\cC_\gamma:
\begin{cases}
  0<x< L_1\\
  \delta_2< y < L_2+\delta_2
\end{cases}&
\cC_\delta:
\begin{cases}
  \delta_1< x < L_1+\delta_1\\
  \delta_2< y < L_2+\delta_2
\end{cases}
\end{matrix}
\end{equation}

In each chart we define a local magnetic potential by 
\begin{equation}
  \label{eq:2}
\forall i:  \bs{\cA}_i = (-\half B y, \half B x)
\end{equation}
(remember we use units where $e/c=1$). All local potentials are
defined in the same way, but their \emph{values} are
different. Within the overlaps of the local charts we easily find the
transition functions realizing the gauge transformations from one
description to another. For instance, the chart $\beta$ overlaps
$\alpha$ in two distinct regions,
$\cI^{(1)}_{\alpha\beta}=\{\delta_1<x_\alpha=x_\beta<L_1\}$ and
$\cI^{(2)}_{\alpha\beta}=\{0<x_\alpha<\delta_1,
L_1<x_\beta<L_1+\delta_1\}$.  In the overlap $\cI^{(1)}_{\alpha\beta}$
the value of the potentials coincide, while in $\cI^{(2)}_{\alpha\beta}$
we have 
\begin{equation}
  \begin{split}
  \bs{\cA}_\beta &=  \bs{\cA}_\alpha + (0, \half BL_1) \\
&= \bs{\cA}_\alpha + \nabla \chi_{\alpha\beta}
  \end{split}
\end{equation}
with $\chi_{\alpha\beta} = \half BL_1y$.  The other transition
functions are determined similarly.  For instance in
$\cI^{(2)}_{\alpha\gamma}=\{0<y_\alpha<\delta_2,
L_2<y_\gamma<L_2+\delta_2\}$ it holds
\begin{equation}
  \begin{split}
  \bs{\cA}_\gamma &=  \bs{\cA}_\alpha + (- \half BL_2, 0) \\
&= \bs{\cA}_\alpha + \nabla \chi_{\alpha\gamma}
  \end{split}
\end{equation}
with $\chi_{\alpha\gamma} = - \half BL_2x$.  

Now, to build the Hamiltonian operator, which is {\sl formally\/}
given by the usual minimal coupling, one has to establish the
transition functions proper to the local wave functions. As it is
well-known these are obtained by exponentiating the transition
functions, i.e.
\begin{equation}
  \label{eq:6}
  \psi_\beta(x,y) =   e^{i\chi^{(j)}_{\alpha\beta}}\psi_\alpha(x,y)\qquad  {\rm on}\; \cI^{(j)}_{\alpha\beta}
\end{equation}
Now take a sequence of points $s_1$ converging to $(L_1,y)$ from the
left and a second sequence $s_2$ converging from the right to the same
point. On $s_1$ we have $\psi_\alpha=\psi_\beta \to
\psi_\alpha(L_1,y)$; on $s_2$ we have $\psi_\beta \to
\psi_\alpha(0,y)\exp\{\half i\, B L_1\, y\}$. By continuity of $\psi_\beta$ we get a condition on $\psi_\alpha$ namely 
\begin{equation}
  \label{eq:BC1}
  \psi_\alpha(L_1,y) = e^{\thalf i\, B L_1\, y}\,\psi_\alpha(0,y)\,.
\end{equation}
By a similar argument we find a second condition 
\begin{equation}
  \label{eq:BC2}
  \psi_\alpha(x,L_2) = e^{-\thalf i \,B L_2\, x}\,\psi_\alpha(x,0)\,.
\end{equation}
At this point we are allowed to work on a single local chart (let's choose $\cC_\alpha$) 
and the Hamiltonian is defined by
\begin{equation}
  \label{eq:Ham}
H =   \half(-i\partial_x+ \half B\,y)^2\,+ 
  \half(-i\partial_y- \half B\,x)^2 + V(x,y)
\end{equation}
on a domain of differentiable functions satisfying
Eq.s(\ref{eq:BC1},\ref{eq:BC2}) as boundary conditions.  Notice
that the b.c. are only consistent if Dirac's condition is
satisfied. To see this, compute $\psi(L_1,L_2)$ by applying the
b.c. in two different orders:
\begin{eqnarray}
  \label{eq:3}
  \psi(L_1,L_2) &=& \psi(0,L_2)\,e^{\thalf i B L_1L_2} =  \psi(0,0)\,e^{\thalf i B L_1L_2}  \\
  \psi(L_1,L_2) &=& \psi(L_1,0)\,e^{-\thalf i B L_1L_2} =  \psi(0,0)\,e^{-\thalf i B L_1L_2}  
\end{eqnarray}
hence $\exp\{i\,B\,L_1\,L_2\} =1$.  All this is well--known, but it
was recalled here to introduce the main idea behind the algorithm we
describe in the next section.

\section{The algorithm}\label{sec:algo}
A simple code, based on a discrete approximation of partial
derivatives, is easily produced; the twisted boundary conditions
Eq.s~(\ref{eq:BC1},~\ref{eq:BC2}) are implemented without
difficulty. However this methods has serious limitations in attaining
good accuracies. A test run with $B=2\pi$, $L_1=L_2=2$ performed with
a $64\times64$ grid in configuration space yields the low energy
spectrum (first 20 eigenvalues) with an average error of $1.5\%$.  In
particular the first four eigenvalues, which should coincide with
$\pi$, turn out to be $\pi\times(0.9997, 1.0082, 1.0082, 1.0419)$. With a finer
mesh ($128\times 128$) the error improves ($0.5\% $) but the computing
time grows considerably (from 25 sec to $\approx 400$ sec).  This fact
encourages to design an algorithm with a better accuracy on partial
derivatives. This is achieved by using a ``spectral method'' based on
the Fourier transform.

\subsection{The spectral method}
A very accurate representation of partial derivatives can be obtained
by using Fourier transform, in one of its efficient implementations as
a numerical code; we shall use FFTW
\cite{FFTW}, which is now included in
Matlab. However, Fourier transform assumes a periodic wave-function,
which is not the case with our problem.  The way out is to apply FFT
separately along $x$ and $y$; the $x$ transform is applied to the
function $\phi=\exp\{-\half iB\,x\,y\}\,\psi$, which turns out to be
periodic in $x$ with period $L_1$. The minimal coupling is then
recovered by realizing that
\begin{equation}
(  -i\partial_x + \half B\,y)\,\psi \equiv e^{\half i Bxy}(-i\partial_x\phi) + B\,y\,\psi\;.
\end{equation}
Now the partial derivative can be computed in $x-$Fourier
space. Similarly $\phi=\exp\{\half iB\,x\,y\}\,\psi$ is periodic in $y$ with the right period, and we may compute
\begin{equation}
(  -i\partial_y - \half B\,x)\,\psi \equiv e^{-\half i Bxy}(-i\partial_y\phi) - B\,x\,\psi\;.
\end{equation}
The idea is used to compute with high accuracy the action of the
Hamiltonian on any function satisfying the twisted b.c.; this is then
used as the unique piece of information needed by the Arnoldi
algorithm to get the spectrum. We also have to choose an initial
vector, if do not feel easy about a {\sl random} initial vector. A
function satisfying the boundary conditions can be constructed as
follows. Choose any $\psi_0(x,y)$, e.g. a Gaussian centered in the
middle of the rectangle of sides $L_1, L_2$. Let $BL_1L_2=2\pi N$; 
then the following equation defines  a ``good'' wave-function:
\begin{equation}
  \psi(x,y) = \sum_{n_1,n_2} (-1)^{n_1\,n_2\,N}\,e^{ \half i B L_2\,x - \half i B L_1\,y  }\, \psi_0(x+n_1L_1, y+n_2L_2)
\end{equation}
The series can be truncated if $\psi_0$ is a Gaussian with a 
width small with respect to $L_i$.

These are the ingredients which can be used to make a call to Matlab's
routine {\tt eigs}\footnote{The Matlab code can be found at the
  author's web site {\tt
    http://www.fis.unipr.it/${\scriptstyle\sim}$enrico.onofri}.},
which provides a very friendly interface to the Arnoldi package {\tt
  Arpack}\cite{ARP}.  The result is rather spectacular as we report
next.

\subsection{Test runs and error estimates}
We apply the algorithm to a grid $n\times n$, starting with very
coarse grids.  In Tab.1 we report the average error and the timings to
compute the first 20 eigenvalues with the same data as before.

\begin{table}[ht]
   \begin{center}
     \begin{tabular}{|c||c|c|}\hline
       n &  Relative Error & Time (sec)\\\hline
       8 & $1.8\times 10^{-2}$        & 0.15\\ \hline
       10 & $2.5\times10^{-3}$       & 0.25\\ \hline
       12 & $5.3\times 10^{-7}$       & 0.35\\ \hline
       16& $1.0\times 10^{-13}$    & 0.60\\\hline
       24 & $1.8\times 10^{-13}$       & 1.35\\ \hline
       32& $2.4\times10^{-13}$ & 2.85\\\hline
       64& $6.0\times10^{-13}$ & 24.7\\\hline
     \end{tabular}\\[.25em]
\caption{}
   \end{center}
 \end{table}  

 As we see, the algorithm reproduces the correct spectrum (including
 degeneracy) already at very low $n$. The relative error saturates
 around $10^{-12}$ which seems to be inherent to the Arnoldi algorithm
 as implemented in Matlab (routine {\tt eigs}). 

\begin{figure}[ht]
  \centering
  \includegraphics[width=.65\textwidth]{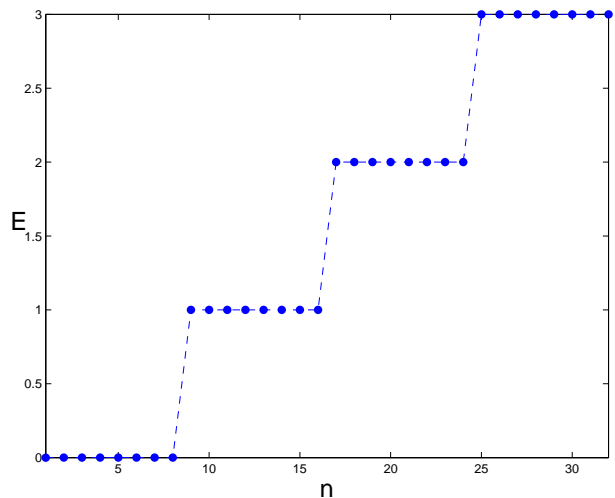}        
  \caption{The Landau levels with $N=\tfrac1{2\pi} BL_1L_2=8$, in units of
    the Larmor frequency.}
  \label{fig.1}
\end{figure}
\noindent
In Fig.1 we see a typical spectrum obtained with the algorithm. The
degeneracy of the eigenvalues is within $10^{-12}$, obtained with a
$32\times 32$ grid.

 Let us notice that if we plug a value of $B$ which does not respect
 Dirac'c condition, the degeneracy is broken; this fact can be
 interpreted as due to the fact that there is a spurious singular
 contribution to the magnetic field at the boundary of the local chart
 which breaks the original symmetry.

 Another check for accuracy can be performed by adding a potential
 energy $\half\omega^2( x^2+y^2)$, in which case the spectrum is known
 in the limit of large $L_1$ and $L_2$. In the case $B=2, \omega=1$,
 we get the spectrum $E=(n_1+\half)\omega_1+(n_2+\half)\omega_2$ with
 a relative error of $10^{-13}$ on a $64\times64$ grid. The presence
 of a potential energy requires a relatively finer mesh.

\section{Fine structure Landau-Mathieu levels}
Having an algorithm which allows for accurate eigenvalue computations
is like having a microscope with higher resolution power: you can
resolve details which would otherwise be invisible. It came then as a
surprise, using the new algorithm, to discover a structure in Landau
levels when the uniform magnetic field is perturbed by an undulating
additive contribution $B\to B(1+\lambda \sin(2\pi \nu
x/L_1)$. Notice that boundary conditions adapted to this choice of
gauge fields must be reformulated, along the lines of
Sec.~\ref{sec:local-charts-twisted}.  Fig.2 shows the splitting of the
first Landau level which occurs at $\nu=4$.
\begin{figure}[ht]
  \centering
  \includegraphics[width=.75\textwidth]{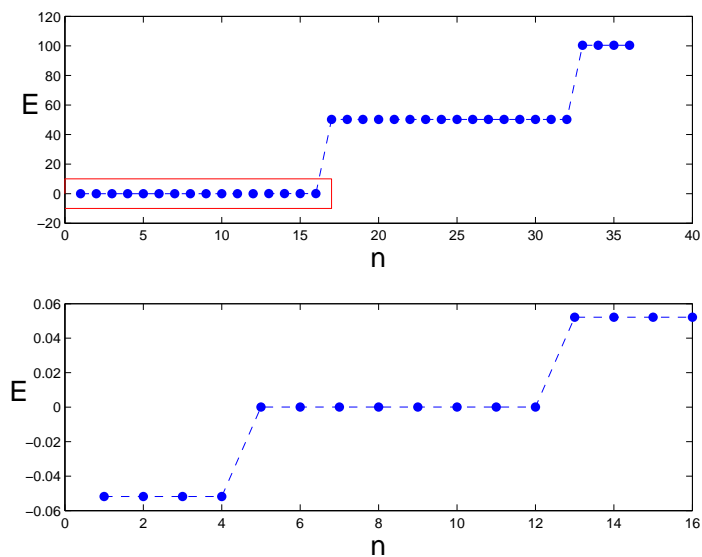}        
  \caption{\small The fine structure pattern of the first Landau level, $\nu=4, N_0=16$. The picture below is a blow--up of the portion of the  rectangle in the picture above. }
  \label{fig.2}
\end{figure}
The pattern is reproduced for other choices of parameters and it looks
numerically very stable and degeneracy within the fine structure
levels is observed numerically at 12 digits precision.(see
Fig.~\ref{fig.3}). There are $N_0=\frac1{2\pi} BL_1L_2$ states in the
first level; these are subdivided in finer sub-levels if $N_0$ is a
multiple of $\nu$: degeneracy is given by the greatest common divisor
${\rm gcd}(N_0, 2\nu)$, hence it is destroyed if $N_0$ and $\nu$ are
relatively prime, but it is left unchanged if $\nu$ is a multiple of
$N_0$.

We also explored the stability of this phenomenon with respect to
deformation of the magnetic field, by keeping its periodicity on the
torus, e.g. by adding a higher harmonic contribution; the pattern of
degeneracy stays the same, only the eigenvalues are shifted (see
Fig. 4).

\begin{figure}[ht]
  \centering
  \includegraphics[width=.65\textwidth]{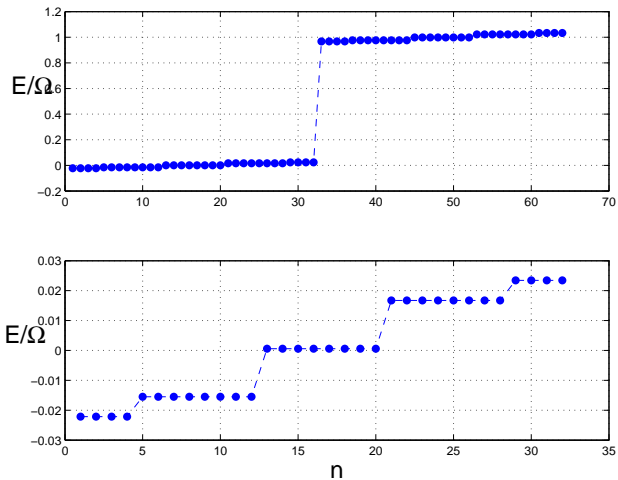}        
  \caption{\small The fine structure pattern of the first Landau
    level for $\nu=4$, $N_0=32$, $\lambda=1/10$.}
  \label{fig.3}
\end{figure}
The finite structure energy gap is not uniform, but a regular pattern emerges looking at sufficiently large $N_0/\nu$. The evidence is that the gaps are approximately reproduced by
\begin{equation}
  \label{eq:7}
  E_{n+1}-E_n \propto \sin(n\pi/N_0)\,,\quad n+\nu/2\equiv 0 \mod(2\nu)\,,
\end{equation}
at least when the degeneracy pattern $\{\nu, 2\nu,
2\nu,...,2\nu,\nu\}$ is realised.  At this level, however, the study
is still preliminary.

\begin{figure}[ht]
  \centering
  \includegraphics[width=.6\textwidth]{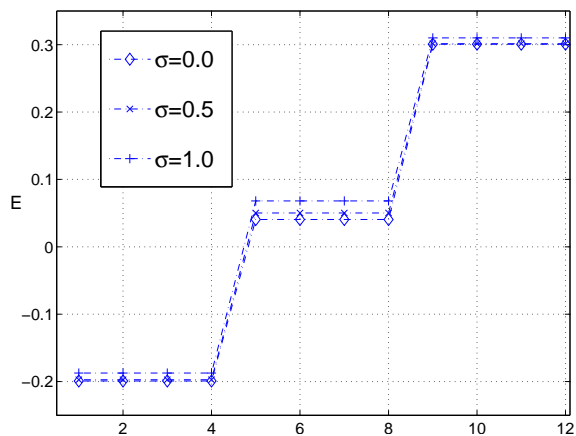}        
  \caption{\small The gaps in the fine structure of the first Landau
    level under a periodic deformation of the magnetic field, $\nu=4,
    N_0=12$, $\sigma$ is the coupling of the higher harmonic.}
  \label{fig.4}
\end{figure}
\section{Concluding remarks}
We presented a spectral algorithm which can compute the energy
spectrum for a scalar particle on the 2-D flat torus, subject to a
transverse magnetic field and any potential energy.  To realize the
algorithm, it is crucial to implement the correct boundary conditions
in order to be able to apply the spectral method based on Fourier
transform. The spectrum is typically obtained to a relative error of
$10^{-12}$ even on rather coarse meshes.  When the field deviates from
uniformity in a sinusoidal way, we find a fine structure in the
splitting of Landau levels with a regular degeneracy pattern. The
problem we considered here originated from the formulation of the
Hamiltonian path integral introduced long ago by J.R.~Klauder
\cite{Klauder}; see also \cite{KlauderEO}

\section*{Acknowledgments}
I would like to warmly thank Professor Andr\'e Neveu and Vladimir
Fateev for the kind hospitality I enjoyed at the LPTA-Montpellier
while this paper has been written. I thank Professor Claudio Destri
for stimulating discussions.  The problem arose in the context of a
Laboratory course at the University of Parma; thanks are due to my
students for providing an efficient stimulus towards the solution.
\bibliographystyle{alpha}

\end{document}